\documentclass[sn-mathphys,Numbered]{sn-jnl}

\usepackage{graphicx}%
\usepackage{multirow}%
\usepackage{amsmath,amssymb,amsfonts}%
\usepackage{amsthm}%
\usepackage{mathrsfs}%
\usepackage[title]{appendix}%
\usepackage{xcolor}%
\usepackage{textcomp}%
\usepackage{manyfoot}%
\usepackage{booktabs}%
\usepackage{algorithm}%
\usepackage{algorithmicx}%
\usepackage{algpseudocode}%
\usepackage{listings}%

\begin{document}

\title[On Galileo’s self-portrait Mentioned by Thomas Salusbury
]{On Galileo’s self-portrait Mentioned by Thomas Salusbury
}


\author*[1]{\fnm{Paolo} \sur{Molaro}}\email{paolo.molaro@inaf.it}

\affil*[1]{\orgdiv{INAF-OATs}, \orgname{Organization}, \orgaddress{\street{Via G.B.Tiepolo 11}, \city{Trieste}, \postcode{34143}, \state{Italia}, \country{}}}


\abstract{An intriguing reference to the existence of a self-portrait by Galileo Galilei is contained in the biography of the scientist by Thomas Salusbury dated ca. 1665, of which only one incomplete and inaccessible copy exists. Galileo grew up in a Renaissance atmosphere, acquiring an artistic touch. He was a musician, a writer and also a painter, as reported by Viviani and documented by his watercolours of the Moon and drawings of solar spots. Recently a new portrait with a remarkable similarity to the portraits of Galileo Galilei by Santi di Tito (1601), Domenico Tintoretto (ca. 1604), and Furini (ca. 1612) has been found and examined using sophisticated face recognition techniques \footnote{R. Srinivasan, C. Rudolph, and A. K. Roy-Chowdhury, {\it Computerized Face Recognition in Renaissance Portrait Art: A quantitative measure for identifying uncertain subjects in ancient portraits}, IEEE Signal Processing Magazine 32, no. 4 (2015): p. 85.}. If the identity could be confirmed, other elements, such as the young age of Galileo or the seam in the canvas revealed by infrared and X-ray analysis, may suggest a possible link with the self-portrait mentioned by Salusbury.}




\maketitle

\section{Galileus Galileus, his life by Thomas Salusbury }\label{sec1}
On 6 June 1660, a few months before the Royal Society was founded at Gresham College, Thomas Salusbury gave the first manuscript of his life’s work to the editor \footnote{Nick Wilding, {\it The Return of Thomas Salusbury's "Life of Galileo" (1664)}, The British Journal for the History o f Science 41, no. 2 (2008): p. 241. }. The whole project was divided into two Tomes, each consisting of two parts and with each part containing several translations and scientific books. The first part of the First Tome contains an English translation of Galileo’s {\it Dialogues} (which at the time was still banned in Italy), Galileo’s letter to the Grand Dukesse Christina Lorena, and a text by Kepler and one by Foscardini. The second part consists mainly of a collection of Benedetto Castelli’s works on the dynamics of fluids. 
The first part of the Second Tome contains Galileo’s Discourses on Two New Sciences, which  turned out to be quite important for Newton’s work –  the first Discourse on the things that move in and upon the water, and other works by Cartesius, Archimedes and Tartalia. The second part contains a work by Torricelli, a book by Salusbury himself and his biography of Galileo, {\it Galileus Galileus, his life, in five books}. This structure suggests that the book was possibly meant to be published separately. The first part of the second Tome, printed in 1665, burnt almost completely during the Great Fire of London of 1666 and only 7 copies are presently known, five of which are in the libraries of the British Museum, Cambridge University, Oxford University and University College of London, with one in a private collection in London. Only one copy of the second part of the second Tome survived the Great Fire, in the private library of the Earl of Macclesfield. We know of its existence since it was quoted by several authors in the early nineteenth century; the final reference to it is found in Drinkwater’s {\it Life of Galileo Galilei}, published in 1829. Afterwards the book disappeared from Macclesfield’s library and only re-appeared two hundred years later in 2006 when it was found by Nick Wilding at Sotheby’s in London, when the library of the Earl of Macclesfield was being auctioned. The book was sold to an anonymous buyer for £153,600 and, regrettably, is not available. Nick Wilding, who is one of the few who saw the book, made a detailed report in his paper ‘The Return of Salusbury’. The copy examined by Wilding shows several annotations and probably is a proof copy of the book. This is the first published biography of Galileo since the one by Viviani in 1654 and was printed in 1717. The index of the book – anticipated in the previous volumes – is provided here in the appendix to show the complexity and richness of information of Salusbury’s work, which is 180 pages long. Very little is known about Thomas Salusbury and the way he collected his sources but we know he travelled through Italy and could speak Italian. We know very little about his life, not even the date of his death, which most probably occurred in early 1666 just before the Great Fire.
Although the book is missing, we know an entire passage from a quote made by Drinkwater in his own biography, published in 1830. On reading this passage in Salusbury’s book, Drinkwater was so impressed that he transcribed the text literally to avoid any misunderstanding. Salusbury seems in the following passage to describe a self-portrait of Galileo: 

{\it He did not contemn the other inferior arts, for he had a good hand in sculpture and carving, but His Particular care was to paint well. By the pencil he described what His telescope discovered; in one he exceeded art, in the other, still. Osorius, the eloquent bishop of Sylya, esteems one piece of Mendoza the wise Spanish minister's felicity, to this, That he was contemporary to Titian, and That by His hand he was drawn in a fair tablet. And Galilaeus, lest he Should want the same good fortune, made so great a progress in this curious art, That he Became His Own Baonarota; and Because there was no other copy worthy of His pencil, drew himself. \footnote{E. J. Drinkwater, Life of Galileo Galilei with Illustrations of the Advancements of Experimental Philosophy (London: printed by William Clowes, 1829), p. 103.}}

Drinkwater added a comment of disbelief at the end: ‘No other author makes the slightest allusion to such a painting; and it Appears Likely That Salusbury Should be mistaken Than That so interesting to portrait Should Entirely lost sight of’ \footnote{ Drinkwater, {\it Life of Galileo Galilei}, p. 103.}.
Drinkwater could be right but the perspective of his epoch is quite different from that of the seventeenth century. In those days it was quite common not to sign or mention paintings and we should not be much surprised if this painting was never mentioned in the short biography of Galileo by Vincenzo Viviani. De’ Nelli could have missed it in absence of clear documentation, as was the case for many things in the first half of Galileo’s life. For instance, we know very little of the other Galileo portraits painted when he was not yet famous, nor of the portrait which was the source of the engravings of Galileo’s image in several of his books.

\begin{figure}[ht]%
\centering
\includegraphics[width=0.9\textwidth]{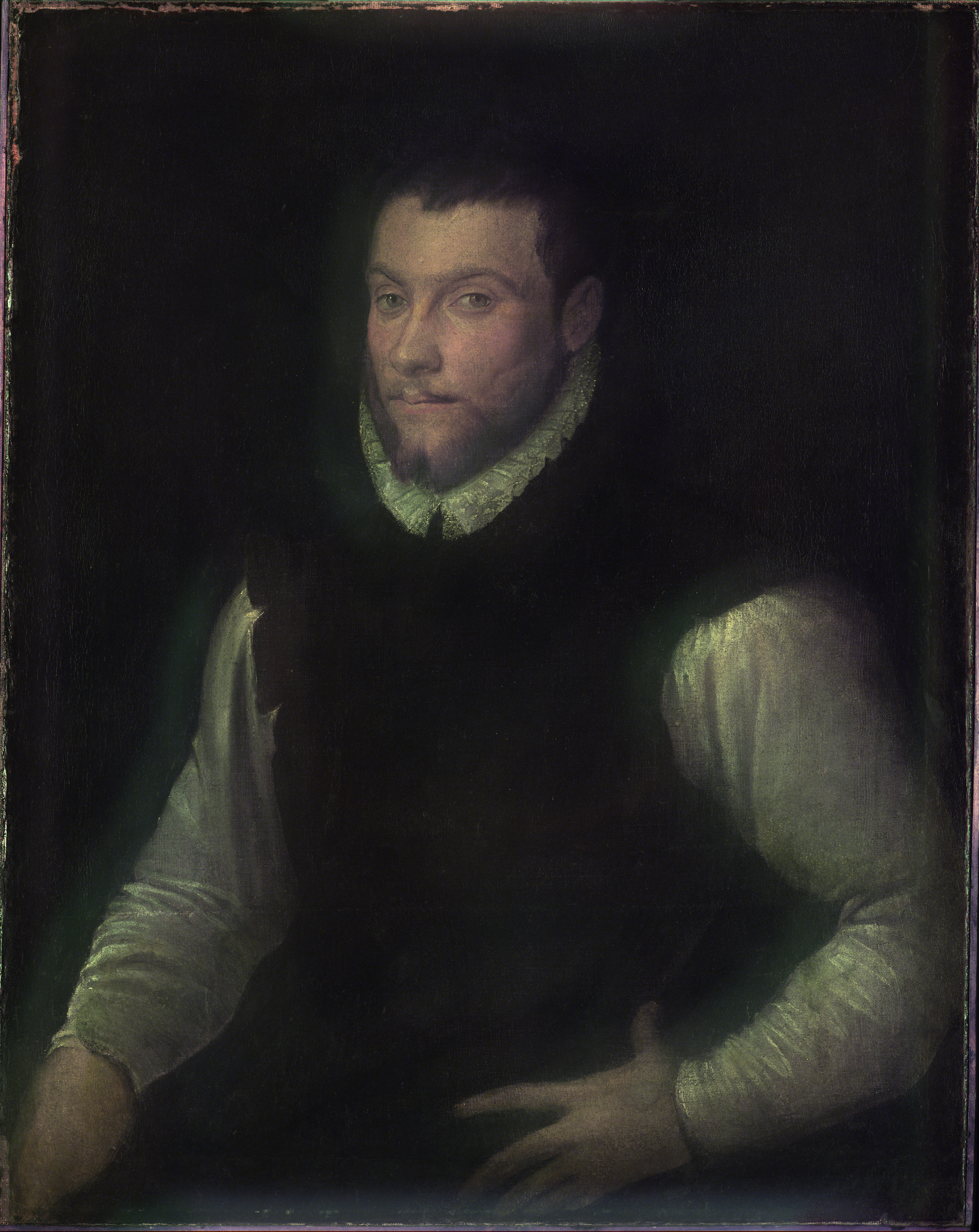}
\caption{Noble gentleman. Is he Galileo?
}\label{fig1}
\end{figure}

\section{The Tuscan Artist}\label{sec2}

Galileo, before becoming the first modern scientist, was a son of the Italian Renaissance. All his biographers agree that he experienced all the arts, from music to literature and painting
\footnote{E. Panofsky, {\it Galileo as a Critic of the Arts. Aesthetic Attitude and Scientific Thought}, Isis 47 (1956): pp. 3–15.}. According to Viviani, Galileo was a talented musician who could compete in playing the lute with his father Vincenzo, one of the most important musicians of the time, and was also an excellent writer, as recognized by Lepardi and Calvino who considered him {\it the greatest Italian writer of all time}. John Milton, who visited Galileo when he was imprisoned in his house in Arcetri, calls him the ‘Tuscan Artist’ in his {\it Paradise Lost}: 

[…] the Moon, whose  Orb\\
Through Optic Glass the TUSCAN Artist views\\
At Ev’ning from the top of FESOLE,\\
Or in VALDARNO, to descry new lands, \\
Rivers or Mountains in her spotty Globe.\footnote{J. Milton, Paradise Lost (London: Samuel Simmons, 1667), 1.297–91.}

Galileo’s interest and talent for painting is carefully documented by Viviani, who wrote:

{\it He greatly enjoyed drawing which he pursued with admirable profit, demonstrating such genius and talent in it that he himself would often say to his friends, that if at the time it had been in his power to decide his own profession, he would certainly have chosen painting. And in truth he had such a natural inclination towards drawing, and over time acquired such an exquisite taste, that his opinions on paintings and drawings were more highly appreciated than those of the leading masters by the masters themselves, such as Cigoli, Bronzino, Passignano and Empoli, and other famous painters of his time, all of whom were great friends of his. Indeed they frequently turned to him for advice about how to organise the narrative, how to arrange the figures, about perspective and colours and all other aspects that contributed to the perfection of painting. They did so because they realised that in this noble art Galileo had the most perfect taste and supernatural grace, such as it was impossible to find in the work of any others, even the masters. So much so, that the illustrious Cigoli, whom Galileo reputed to be the foremost painter of his time, attributed most of the best work that he did to the excellent teachings of Galileo himself, and in particular he felt himself honoured to say that in the matter of perspective, he alone had been his master.}

In 1613 Galileo became a member of Vasari's Florentine Academy of Design where he distinguished himself for his taste in supporting and protecting, {\it inter alia}, the young Artemisa Gentileschi. Disappointingly, as all biographic sources agree, few records survive of Galileo’s artistic ability. Among these are the watercolours he made during his first telescopic observations of the Moon in November 1609. The line of the terminator and the white spots in the dark area need to be drawn very precisely since from their distance Galileo computed the height of the mountains on the Moon. Similarly, the drawing of the solar spots published in the {\it Istoria e Dimostrazioni sulle Macchie Solari} (1613) are extremely accurate in shape and position and allow us to compute the solar activity in that period, the first one ever recorded by man \footnote{Galileo Galilei, {\it Istoria e Dimostrazioni intorno alle Macchie Solari e loro accidenti }(Roma: Appresso Giacomo Mascardi, 1613).} . The shape of the solar spots near the solar limb was essential in the Galileo’s understanding that they were on the solar surface and were not intervening spherical bodies, as proposed by Scheiner. It is worth noting that one drawing is dated 26 June 1612, which is the date of Galileo’s famous letter to Cigoli on the ‘Comparison of Arts’ where, requested by his friend, Galileo compares the relative merits of painting and sculpture and argues that painting is superior because it is more distant to reality \footnote{ Galileo Galilei, {\it Letter to Cigoli, 26 June 1612}, OG, XI, p. 340, n. 713.}. A quite modern view indeed. A few other interesting sketches were found by Horst Bredekamp in Galileo’s notes on waste paper, dating to around 1584, certainly something that Galileo would have never shown to anyone.\footnote{Horst Bredekamp, {\it Gazing hands and blind spots: Galileo as draftsman}, Science in Context  14, no. S1 (2001): p. 153; and Horst Bredekamp, Galilei der Künstler: Der Mond, die Sonne, die Hand (Berlin: Akademie Verlag, 2007).}

\begin{figure}[ht]%
\centering
\includegraphics[width=0.9\textwidth]{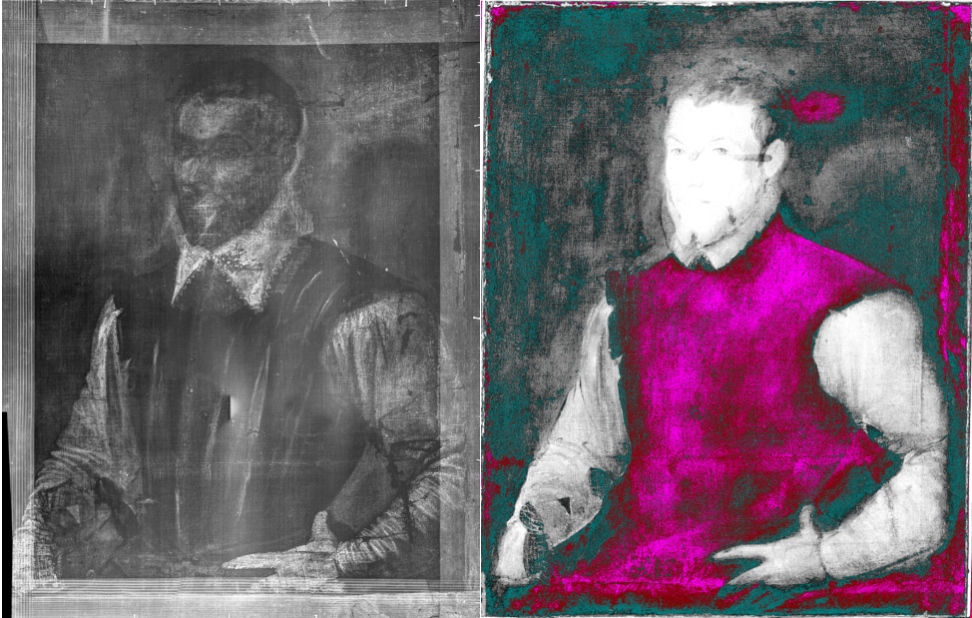}
\caption{ Left: X-ray. Right: IR 2265nm in false colours performed at the INO-CNR in Florence by Raffaella Fontana}\label{fig2}
\end{figure}

\section{A new Galileo portrait?}\label{sec3}
A few years ago an antique painting showing a nobleman has been connected to Galileo \footnote{ Paolo Molaro, {\it Possible Portrait of Galileo Galilei as a young scientist}, Astronomische Nachrichten 333 (2012): p. 186.}. The identification of the gentleman as Galileo is principally based on the resemblance to other portraits of Galileo. The painting shown in Fig. \ref{fig1}  is a half-length portrait of a man dressed in black, turned slightly to the left, but looking directly at the observer. The man has short hair and an incipient red beard. The painting – oil on canvas 87 x 69cm – was classified by the experts as a portrait of the Italian school of the 17th century. The painting has been analysed with UV-Wood, X-ray and Infrared frequencies. The latter was performed at the National Optical Florence INO-CNR by Raffaella Fontana and collaborators using the technique of infrared reflectography. Twelve images were taken between 0.952 and 2.265 microns of the optical band. A representative image is shown in Fig. \ref{fig2}. The pigments are partially transparent to infrared radiation and allow one to see that the preliminary drawing of the face was done with no second thoughts in the execution, while probably the body and hands were sketched up, probably directly with a brush. Interestingly the X-ray image reveals a different and simpler collar. 
A horizontal seam at about 24cm from the bottom edge of the picture combines together two pieces of canvas. At the bottom we can see a re-painting, probably made at a later time and by a different and less skilful hand. X and IR images reveal a different original drawing of the right arm, which holds some objects. These seem to include two rolls of paper tied with two bows and other unidentified tools. This could have been done deliberately to prevent the identification of the sitter and a careful restoration could be very useful in this respect.
The first known portrait of Galileo is that of 1601 by the painter Santi di Tito, as reported by De’ Nelli, who writes:

{\it The most illustrious Italian artists wished to have the honour of painting Galileo’s portrait. Santi di Tito portrayed him in a small painting in 1601 at the age of thirty-eight, not long before he himself passed away. [...] This portrait is the one conserved in my private Library, and I have placed the derived engraving executed by Mr. Giuseppe Calendi at the opening of this History.\footnote{G. B. De' Nelli, Vita e commercio letterario di Galileo Galilei (Losanna: 1793), pp. 872, 873.}
}

The painting by Santi di Tito was then lost, although it may perhaps have been recently rediscovered, as reported by Tognoni in 2013; also see my paper of 2016 on the subject \footnote{F. Tognoni, ed., Le opere di Galileo Galilei: {\it Iconografia galileiana}, Appendice, Vol. 1 (Florence: Giunti Editore, 2013); and F. Tognoni, {\it L’immagine di Galileo:tra iconoteche e biografie illustrate}, in Atti e Memorie dell’Accademia Galileiana di Scienze, Lettere ed Arti già dei Ricovrati e Patavina. Parte III: Memorie della Classe di Scienze Morali, Lettere ed Arti CXXVII (2014–2015). Also see Paolo Molaro, {\it Su ritratto perduto di Galileo ad opera del pittore toscano Santi di Tito}, Giornale di Astronomia 42, no. 1 (2016): pp. 10, 14}.

The very dating of the painting is fairly approximate, having been proposed by Fahie on the basis of the double letters {\it ll} in the legend on the painting, which appears to be an element detected only in documents of that part of the Venetian period \footnote{J. J. Fahie, {\it Memorials of Galileo Galilei (1564-1642), Portraits and Paintings, Medals and Medallions, Busts and Statues, Monuments and Mural Inscriptions} (Leamington and London: Courier Press, 1929), pp. 10–12.}.

\begin{figure}[ht]%
\centering
\includegraphics[width=0.9\textwidth]{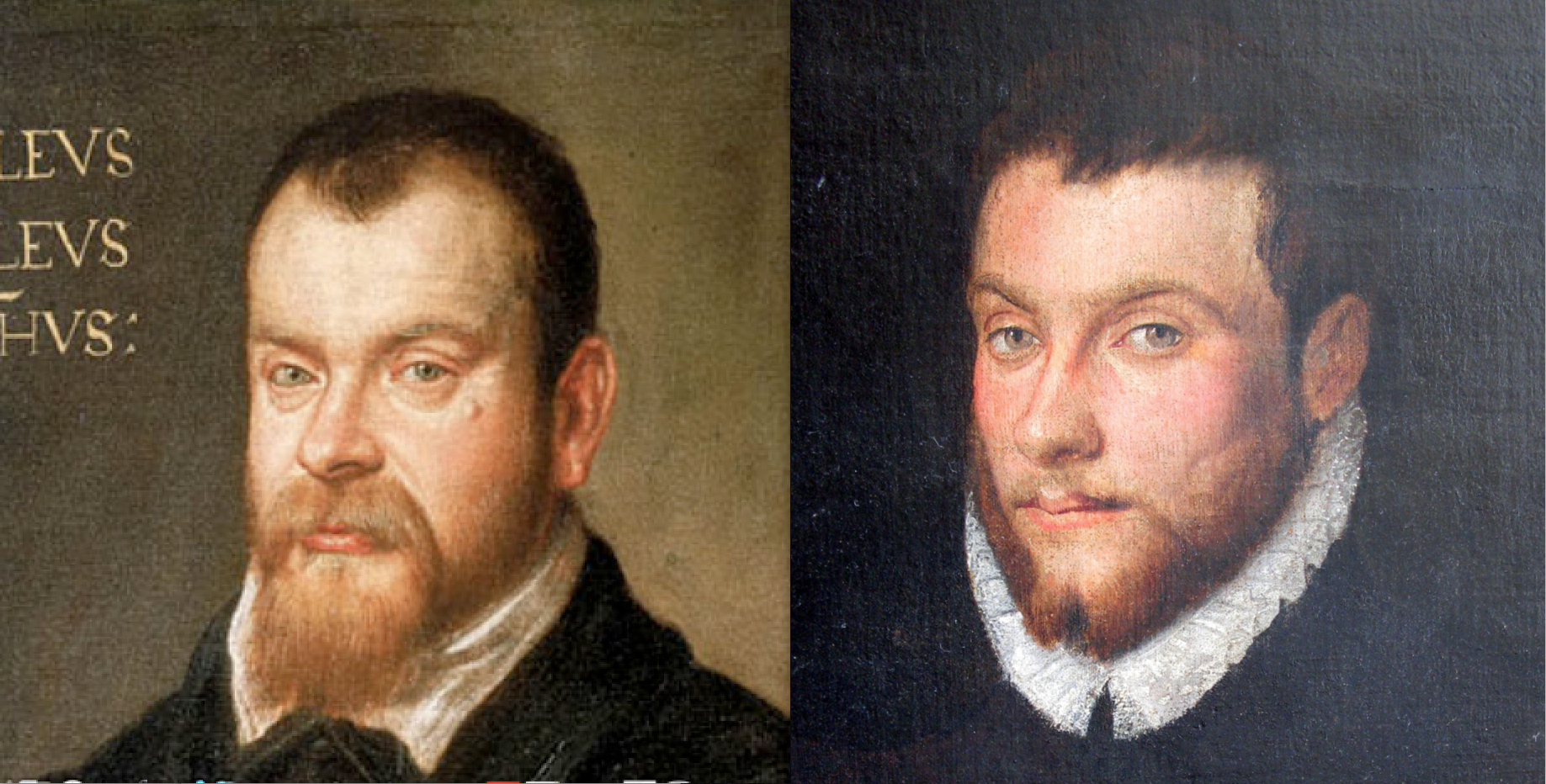}
\caption{ On the left Galileo Galilei by Domenico Tintoretto, and on the right is the anonymous}\label{fig3}
\end{figure}

Another portrait is the one possibly painted by Filippo Furini, known as lo Sciamerone in 1612 and used for the engraving of the frontispiece of the {\it Istoria e dimostrazioni intorno alle macchie solari} of 1613 and in {\it Il Saggiatore} of 1623 \footnote{Galilei, {\it Istoria e dimostrazioni intorno alle macchie solari}, and in Galileo Galilei, {\it Il Saggiatore nel quale con bilancia esquisita e giusta si ponderano le cose contenute nella libra astronomica e filosofica di Lotario Sarsi Sigensano scritto}  (Roma: Appresso Gioacomo Mascardi, 1623).}. The painting is at the Vienna Kunsthistorisches (INV GG 7976) where, however, it bears a different attribution to Tiberio Titi, son of Santi di Tito.
Then there are three identical pencil portraits, dated and signed, which were executed by the papal portraitist Ottavio Leoni in 1624 during Galileo’s visit in Rome to the new Pope, Urban VIII. Of the same year is the oil painting on a plate of silver attributed to the painter Domenico Passignani, now part of a private collection in Helsinki. The two most famous portraits of Galileo which are always mentioned in the various biographies are those by the Flemish painter Justus Susterman, court painter of the Medici, dated 1636 and 1640 and now conserved in Florence in the Palatine Gallery and in the Uffizi. These show Galileo late in life, and represent the paradigm of the Galilean icon. Copies of them are in the Bodleian libraries, sent by Viviani in 1661.
The identification of the gentleman shown in Fig. \ref{fig1}  as Galileo is based on the resemblance to the portrait by Tintoretto and Santi di Tito, obviously taking into consideration the different age. It is true that the characteristic mole above the left cheekbone is missing, but this might only have appeared later. Detail of the face is shown in Fig. \ref{fig3}, where the particular of the face is set in direct comparison with Tintoretto’s portrait. In the Tintoretto the face appears slightly thinner, but the essential proportions are the same. The most significant difference concerns the nose, which is slightly shorter in Tintoretto’s portrait, while the shape and the blue colour of the eyes are extremely similar. The painting is also very similar to the engraving by Giuseppe Calendi. An overlay shows that the shape of the head coincides perfectly, while there is a slight difference in the shape of the tip of the nose and the lobe of the left ear, which is missing in the painting where it is perhaps covered by the beard. 
The painting was recently studied within the framework of the project FACES (Faces, Art, and Computerized Evaluation Systems) coordinated by Conrad Rudolph of the Department of the History of Art, University of California, Riverside, in which the most sophisticated face recognition technology is applied to works of art. Their conclusions regarding the portraits of Anne Boleyn received extensive coverage in the media. Indeed, they rejected the only portrait believed to be of Anne Boleyn, but found a new identification of Anne in a portrait previously believed to be of Jane Seymour. When tested against eight other known portraits of Galileo Galilei, they found consistently decreasing match scores for the three paintings chronologically closest in age (1601–ca. 1612), ‘no decisions’ for the two dated 1624, and non-matches for the final three done when Galileo was old \footnote{ C. Rudolph et al., 2016 in press; Srinivasan et al., {\it Computerized Face Recognition in Renaissance Portrait Art}, IEEE signal processing magazine 32,4 p. 85.}. More specifically, the matches with the portraits by Santi di Tito (via Calendi), Tintoretto and the Furini have maximum probability (probability 0.724, 0.718, 0.708 respectively, with the peak of probability of the procedure at 0.725), whereas they are ambiguous with the Leoni and the Passignano (0.672, 0.650) and are not consistent with those of Susterman (0.639 and 0.590). The latter has to be attributed to the process of ageing that altered the physiognomy of the scientist’s lineaments.
It should be noted that this analysis does not consider colour. Indeed, elements such as the colour of the eyes and hair and of the flesh tones which play a significant role in the processes of human recognition, are not taken into consideration in the FACES procedure. This confirms that the anonymous figure portrayed displays a remarkable similarity to Galileo, even though – in the absence of certain documentation – it is not possible to distinguish between Galileo and a perfectly identical double. If he is Galileo, he must have been aged between 20 and 25, which corresponds to the period during which he became a professor at Pisa or in the early years of his teaching in Padua. The portrait offers us a unique image of the young Galileo, with his intense and captivating gaze, fully aware of his own capacities and almost as if he had a premonition of the important role he was to play in the history of science.
The author of the painting could have been one of the many painters belonging to the circle of his artistic acquaintances, and may have been executed as an exercise. There is also documentation of other portraits executed by the painters of the time which are now lost. For example the one by Ludovico Ciardi –  known as Cigoli – who had been Galileo’s friend since the time they both took lessons of perspective from the mathematician Ostilio Ricci and who portrayed the physical Moon of the Sidereus in the fresco in Santa Maria Maggiore in Rome. The reference to the portrait by Cigoli is contained in a letter from Luca Valerio to Galileo dated 4 April 1609 and is probably too late to be the one we are considering here \footnote{ Luca Valerio, {\it Letter to Galileo of 4 April 1609}, the National Library in Florence, Mss. Gal., P. VI, T. VII car. 93 (autograph), 1609.}
However, since at that time Galileo was anything but famous, and that portraits were painted only for figures of a certain status, the hypothesis of the self-portrait executed by Galileo in his youth, mentioned by Thomas Salusbury, might not be entirely implausible. Certainly it is possible that Salusbury is wrong, as Drinkwater concludes, but it should be not too surprising that no mention to this self-portrait is present in the earliest biographies of Galileo. Viviani’s is very concise and De’ Nelli’s appears very late, when many sources were compromised. These biographies or later ones, including that of Drinkwater, lack references to other portraits of Galileo. Moreover, many famous portraits of Galileo are not signed and can hardly be dated, following the customs of the time. It should not be a surprise that a work probably made in his youth was overlooked. We recall that the whole of Galileo’s life when he was in Padova is very little documented.
In the sixteenth century portraits were commissioned only by important people, but this contrasts with this painting, made in a {\it cheap} way, which shows a degree of economic difficulty of the author or of whoever commissioned the work. Thus, the possibility that Galileo made a self-portrait is an intriguing one which is certainly unlikely but that cannot be excluded {\it a priori}.

\bmhead{Acknowledgments}
It is a pleasure to thank Federico Tognoni, Pier Andrea Dandò, Franco Zotto, Raffaella Fontana, Marco Barucci, Ferruccio Petrucci, Conrad Rudolph, Paolo di Marcantonio and Francesco Palla.

\begin{appendices}

\section{Index of the Thomas Salusbury’s Galileus Galileus, his life, in five books. }\label{secA1}

Index of the Thomas Salusbury’s Galileus Galileus, his life, in five books. The book belongs to an anonymous private collector. We strongly pray its owner in honour of the science of making it accessible.\\

III. GALILEUS GALILEUS, his LIFE: in Five BOOKS\\

BOOK I. Containing Five Chapters.\\
 Chap.
 1. His Country.\\
  Chap.		2. His Parents and Extraction.\\
 Chap.		3. His time of Birth.\\
  Chap.		4. His first Education.\\
 Chap.	 	5. His Masters.\\
 
BOOK II. Containing Three Chapters.\\
 Chap. 1. His judgment in several Learnings.\\
  Chap.		2. His Opinions and Doctrine.\\
 Chap.		 3. His Auditors and Scholars.\\
 
BOOK III. Containing Four Chapters.\\
 Chap. 1. His behaviour in Civil Affairs.\\
  Chap.		2. His manner of Living.\\
  Chap.		3. His morall Virtues.\\
  Chap.		4. His misfortunes and troubles.\\
  
BOOK IV. Containing Four Chapters.\\
 Chap. 1. His person described.\\
  Chap.		2. His Will and Death.\\
  Chap.		3. His Inventions.\\
  Chap.		4. His Writings.\\
 Chap. 		5. His Dialogues of the Systeme in particular, containing Nine Sections. \\				
  Section   1. Of Astronomy in General; its Definition, Praise, Original.\\
  Section 		  		  2. Of Astronomers: a Chronological Catalogue of the most famous of them.\\
 Section 		  		  3. Of the Doctrine of the Earths Mobility, \& its Antiquity, and 						Progresse from Pythagoras to the time of Copernicus.\\
  Section 				  4. Of the Followers of Copernicus, unto the time of Galileus.\\
  Section 				  5. Of the severall Systemes amongst Astronomers.\\
 Section 				  6. Of the Allegations against the Copern. Systeme, in 77 Arguments 	\\				taken out of Ricciolo, with Answers to them.\\
Section 				  7. Of the Allegations for the Copern. Systeme in so Arguments.\\
Section 				  8. Of the Scriptures Authorities produced against and for the 							Earths mobility.
Section 				  9. The Conclusion of the whole Chapter.\\

BOOK V. Containing Four Chapters.\\
 Chap. 1. His Patrons, Friends, and Emulators.\\
Chap.		2. Authors judgments of him.\\
Chap.		3. Authors that have writ for, or against him.\\
Chap.		4. A Conclusion in certain Reflections upon his whole Life.\\

\end{appendices}



\end{document}